\documentclass[aps,prl,twocolumn,showpacs,superscriptaddress,groupedaddress]{revtex4}  
\usepackage{graphicx}  
\usepackage{dcolumn}   
\usepackage{bm}        
\usepackage{amssymb}   

\begin{document}

\title{Probing in-medium spin-orbit interaction with intermediate-energy heavy-ion collisions}
\author{Jun Xu}\email{xujun@sinap.ac.cn}
\affiliation{Shanghai Institute of Applied Physics, Chinese Academy
of Sciences, Shanghai 201800, China}
\affiliation{Department of
Physics and Astronomy, Texas A$\&$M University-Commerce, Commerce,
TX 75429-3011, USA}
\author{Bao-An Li}\email{Bao-An.Li@tamuc.edu}
\affiliation{Department of Physics and Astronomy, Texas A$\&$M
University-Commerce, Commerce, TX 75429-3011, USA}
\affiliation{Department of Applied Physics, Xi'an Jiao Tong
University, Xi'an 710049, China}

\begin{abstract}

Incorporating for the first time both the spin and isospin degrees
of freedom explicitly in transport model simulations of
intermediate-energy heavy-ion collisions, we observe that a local
spin polarization appears during collision process. Most
interestingly, it is found that the nucleon spin up-down
differential transverse flow is a sensitive probe of the spin-orbit
interaction, providing a novel approach to probe both the density
and isospin dependence of the in-medium spin-orbit coupling that is
important for understanding the structure of rare isotopes and
synthesis of superheavy elements.

\end{abstract}

\pacs{25.70.-z, 
      24.10.Lx, 
      13.88.+e  
      }

\maketitle The spin-orbit coupling is common to the motion of many
objects in nature from quarks, nucleons, and electrons to planets
and stars. In nuclear physics, it has been very well known for a
long time that the spin-orbit coupling is crucial for understanding
the structure of finite nuclei, such as the magic
numbers~\cite{Goe49,Hax49}. However, many interesting questions
regarding the in-medium spin-orbit coupling, especially its density
and isospin dependence, remain unresolved although in free space it
has been well determined from nucleon-nucleon (NN) scattering
data~\cite{Wir95}. Indeed, several phenomena observed or predicted
in studying properties of nuclear structures have been providing us
some useful information about the in-medium spin-orbit interaction.
For instance, the kink of the charge radii of lead isotopes can only
be explained by introducing a weakly isospin-dependent spin-orbit
coupling~\cite{Sha95,Rei95}. Moreover, strong experimental evidences
of a decreasing spin-orbit coupling strength with increasing neutron
excess were reported~\cite{Sch04,Gau06}. Furthermore, new
experiments are currently being carried out at several laboratories
to explore the density and isospin dependence of the spin-orbit
coupling by comparing energy splittings of certain orbits in the
so-called ``bubble" nuclei~\cite{Tod04,Gra09} with those in normal
nuclei~\cite{Sorlin12}. The knowledge of the in-medium spin-orbit
interaction is useful for understanding properties of drip-line
nuclei~\cite{Lal98}, the astrophysical r-process~\cite{Che95}, and
the location of stability island for superheavy
elements~\cite{Ben99,Mor08}.

While effects of the spin-orbit interaction on nuclear structure
have been studied extensively, very little is known about its
effects in heavy-ion collisions. Within the time-dependent
Hartree-Fock (TDHF) calculations the nucleon spin-orbit interaction
was found to affect the fusion threshold energy~\cite{Uma86} and
lead to a local spin polarization~\cite{Mar06,Iwa11}. In non-central
relativistic heavy-ion collisions at 200 GeV/nucleon the partonic
spin-orbit coupling was found to lead to a global quark spin
polarization~\cite{Lia05}. However, to our best knowledge, no study
on effects of the spin-orbit interaction in intermediate-energy
heavy-ion collisions has been carried out yet. On the other hand,
several facilities for spin-polarized beams have been developed for
about twenty years. It has already been shown that spin-polarized
projectile fragments in peripheral collisions are measurable through
the angular distribution of $\gamma$ or $\beta$ decays at both GSI
and RIKEN~\cite{Sch94,Ich12}. In addition, at AGS and RHIC energies,
people can already obtain the spin-flip probability and distinguish
spin-up and spin-down nucleons in elastic pp or pA collisions by
measuring the analyzing power~\cite{Toj02}. One thus expects that
spin-related experimental observables in intermediate-energy
heavy-ion collisions can be measured in the near future, if they are
indeed helpful for enriching our knowledge about the poorly known
in-medium spin-orbit interaction. In this Letter, within a newly
developed spin-isospin dependent transport model, we show that the
nucleon spin up-down differential transverse flow in heavy-ion
collisions at intermediate energies is a sensitive probe of the
density and isospin dependence of the in-medium nucleon spin-orbit
interaction.

Starting from the following effective nucleon spin-orbit interaction~\cite{Vau72}\label{vso}
\begin{equation}
V_{so} = i W_0 (\vec{\sigma}_1+\vec{\sigma}_2) \cdot \vec{k} \times
\delta(\vec{r}_1-\vec{r}_2) \vec{k}^\prime,
\end{equation}
where $W_0$ is the strength of the spin-orbit coupling,
$\vec{\sigma}_{1(2)}$ is the Pauli matrix,
$\vec{k}=(\vec{p}_1-\vec{p}_2)/2$ is the relative momentum operator
acting on the right with $\vec{p}=-i\nabla$, and $\vec{k}^\prime$ is
the complex conjugate of $\vec{k}$, the single-particle Hamiltonian
in nuclear matter can be written as
\begin{equation}\label{h}
h_q = \frac{p^2}{2m} + U_q + U_q^s + U_q^{so}
\end{equation}
where $q=n$ or $p$, $U_q$ and $U_q^s$  are the central bulk and spin
potential, respectively, and $U_q^{so}$ is the spin-orbit potential.
In this work, we use a momentum-independent $U_q$ leading to an
incompressibility of $K_0=230$ MeV for symmetric nuclear matter, a
symmetry energy $E_{sym}=30$ MeV and its density slope $L=60$ MeV at
saturation density $\rho_0=0.16$ fm$^{-3}$ similar to the
parameterization of a modified Skyrme-like interaction~\cite{Che10}.
The $U_q^s$ and $U_q^{so}$ can be expressed respectively as,
\begin{eqnarray}
U_q^s &=& -\frac{W_0}{2} [\nabla \cdot (\vec{J} + \vec{J}_q) ] -
\frac{W_0}{2}\vec{p} \cdot [\nabla \times (\vec{s} +
\vec{s}_q) ] \nonumber \\
&-& \frac{W_0}{2} \vec{\sigma} \cdot [\nabla \times (\vec{j} +
\vec{j}_q) ], \label{us}\\
U_q^{so} &=& \frac{W_0}{2}(\nabla \rho + \nabla \rho_q) \cdot
(\vec{p} \times \vec{\sigma}), \label{uso}
\end{eqnarray}
where $\vec{J}$, $\vec{s}$, $\vec{j}$, and $\rho$ are the nucleon
spin-current, spin, momentum, and number densities, respectively. We
notice here that the second and third terms in Eq.~(\ref{us}) are
the time-odd contributions~\cite{Eng75} suppressing the first term
in Eq. (\ref{us}) and the term in Eq. (\ref{uso}), respectively, and
neglecting them would break the Galilean invariance and induce a
spurious spin excitation~\cite{Mar06}. Taking into account both the
density~\cite{Pea94} and isospin dependence~\cite{Sha95} of the
spin-orbit interaction, the $U_q^s$ and $U_q^{so}$ can be generally
written as
\begin{eqnarray}
U_q^s&=&-\frac{W_0^\star(\rho)}{2} [\nabla \cdot (a\vec{J}_q +
b\vec{J}_{q^\prime}) ] - \frac{W_0^\star(\rho)}{2}\vec{p}
\cdot[\nabla \nonumber \\
&\times& (a\vec{s}_q+ b\vec{s}_{q^\prime}) ]
-\frac{W_0^\star(\rho)}{2} \vec{\sigma} \cdot [\nabla \times
(a\vec{j}_q +
b\vec{j}_{q^\prime}) ], \label{usg}\\
U_q^{so} &=& \frac{W_0^\star(\rho)}{2}(a\nabla \rho_q + b\nabla
\rho_{q^\prime}) \cdot (\vec{p} \times \vec{\sigma}). (q \ne
q^\prime) \label{usog}
\end{eqnarray}
In the above, $W_0^\star(\rho) = W_0(\rho/\rho_0)^\gamma$ represents
the density-dependence of the spin-orbit coupling. Different
combinations of $\gamma$, $a$, and $b$ can be used to mimic various
density and isospin dependences of the in-medium spin-orbit
interaction while preserving the Galilean invariance. We
notice that with $\gamma=0$, $a=2$, and $b=1$ Eqs.~(\ref{usg}) and
(\ref{usog}) reduce to Eqs.~(\ref{us}) and (\ref{uso}), while equal
values for $a$ and $b$, and a nonzero value for $\gamma$ were
predicted within a relativistic mean-field model~\cite{Rei95}.
Neglecting the density dependence of $W_0^\star$, the spin-orbit
coupling constant $W_0$ ranges from about $80$ MeVfm$^5$ to $150$
MeVfm$^5$~\cite{Les07,Zal08,Ben09}, while the values of $\gamma$,
$a$, and $b$ are still under hot debate.

To model the spin-isospin dynamics in heavy-ion collisions at
intermediate energies, we incorporate explicitly the spin degree of
freedom and the spin-related potentials in a previously developed
isospin-dependent Boltzmann-Uehling-Uhlenbeck (BUU) transport model,
see, e.g., Refs.~\cite{Li97,LCK}. The new model is dubbed SI-BUU12.
To our best knowledge, the spin degree of freedom was never
considered before in any of the existing transport models for
heavy-ion reactions at intermediate energies since the emphasis of
the community has been on extracting information about the Equation
of State (EOS) of symmetric nuclear matter, density dependence of
nuclear symmetry energy, and in-medium NN scattering cross sections
using spin-averaged experimental observables. In the SI-BUU12 model,
$\rho$, $\vec{s}$, $\vec{J}$, and $\vec{j}$ are all calculated by
using the test particle method~\cite{Won82,Ber88}. The equations of
motion in the presence of the spin-orbit interaction are
\begin{eqnarray}
\frac{d\vec{r}}{dt} &=& \frac{\vec{p}}{m} +
\frac{W_0^\star(\rho)}{2} \vec{\sigma} \times (a\nabla \rho_q +
b\nabla \rho_{q^\prime}) \nonumber \\
&-& \frac{W_0^\star(\rho)}{2} \nabla \times (a\vec{s}_q + b\vec{s}_{q^\prime}), \label{rt}\\
\frac{d\vec{p}}{dt} &=& - \nabla U_q - \nabla U_q^s - \nabla
U_q^{so}, \label{pt}
\\
\frac{d\vec{\sigma}}{dt} &=& W_0^\star(\rho)[(a\nabla \rho_q +
b\nabla \rho_{q^\prime}) \times \vec{p}] \times
\vec{\sigma} \nonumber \\
&-&W_0^\star(\rho)[\nabla \times (a\vec{j}_q + b\vec{j}_{q^\prime})]
\times \vec{\sigma}. \label{sigmat}
\end{eqnarray}
In the center-of-mass frame of nucleus-nucleus collisions, two Fermi
spheres of projectile/target nucleons are Lorentz boosted in the
$\pm z$ direction. Note that the $U_q^{so}$ and the last term in
$U_q^s$ are most important among the spin-related potentials, with
the former being the time-even contribution while the latter being
the time-odd contribution. During the collision process, the density
gradient $\nabla \rho$ points mainly along the impact parameter of
the reaction, i.e., the x axis, while the momentum density $\vec{j}$
is mainly located in the reaction plane (x-o-z) and $\nabla \times
\vec{j}$ is thus along the y axis perpendicular to the reaction
plane. Due to the spin-orbit potential, the nucleon spin
$\vec{\sigma}$ tends to be parallel to the direction of $\vec{p}
\times \nabla \rho$ in order to lower the energy of the system. On
the other hand, the time-odd contribution makes the nucleon spin
$\vec{\sigma}$ parallel to $\nabla \times \vec{j}$, which is in the
opposite direction of $\vec{p} \times \nabla \rho$~\cite{Mar06}. The
result of their competition determines the final direction of the
nucleon spin. We will refer in the following a nucleon with its spin
in the +y (-y) direction as a spin-up (spin-down) nucleon. During
heavy-ion collisions at intermediate energies with different
combinations of targets, projectiles, and impact parameters,
dynamical systems of nucleons with different density gradients,
momentum currents, and the isospin asymmetries are formed. These
reactions thus provide a useful tool for investigating the density
and isospin dependence of nucleon spin-orbit coupling. Of course, to
realize this goal the first challenge is to find sensitive
experimental observables. Transport models have been very successful
in both extracting reliable information about the nuclear EOS and
predicting new phenomenon that have later been experimentally
confirmed, see, e.g., Refs. \cite{LCK,Dan02a}. We are confident that
the SI-BUU12 model has similar predictive powers for studying spin
physics with intermediate-energy heavy-ion collisions.

Nucleon-nucleus scattering experiments have shown that nucleons may
flip their spins after NN scatterings due to spin-related nuclear
interactions, see, e.g., Ref.~\cite{Ohl72} for a review. Although
not well determined yet, the spin-flip probability for in-medium NN
scatterings is known to be appreciable, depending on the collision
energy and the momentum transfer~\cite{Lov81}. In the present work
we will test different options of setting spins, including
randomizing, flipping, or keeping spins unchanged after each NN
scattering to study effects of different spin-flip probabilities on
spin-sensitive observables. In addition, a spin- and isospin-dependent Pauli
blocking is introduced in the SI-BUU12 model.

\begin{figure}[ht]
\includegraphics[scale=0.4]{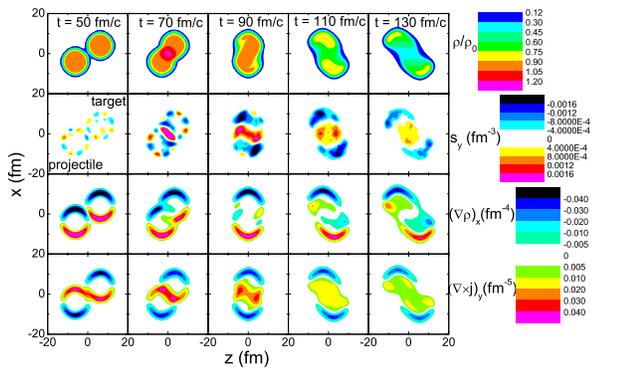}
\caption{(color online) Contours of nucleon reduced density
$\rho/\rho_0$ (first row), y component of spin density $s_y$ (second
row), x component of the density gradient $(\nabla \rho)_x$ (third
row), and y component of the curl of the momentum density $(\nabla
\times \vec{j})_y$ (fourth row) in the reaction plane at different
stages in Au+Au collisions at a beam energy of 50
MeV/nucleon and an impact parameter of $\text{b}=8$ fm. 
}\label{den}
\end{figure}

As an illustration of the effects from spin-orbit coupling, it is
interesting to first examine for a typical reaction at intermediate
energies the time evolution of the y component of the spin density
$s_y$, the x component of the density gradient $(\nabla \rho)_x$,
and the y component of the curl of the momentum density $(\nabla
\times \vec{j})_y$ in comparison with the nucleon density contour in
the reaction plane. Shown in Fig.~\ref{den} are these quantities for
Au+Au collisions at a beam energy of 50 MeV/nucleon and an impact
parameter of $\text{b}=8$ fm. For this example, we set $W_0=150$
MeVfm$^5$, $\gamma=0$, $a=2$, $b=1$, and the spins of the colliding
nucleons are randomized after each scattering. Initially we put the
projectile and target nuclei without spin polarization far away, and
there is no spin polarization before they physically meet each other
due to the cancelation of the time-even and time-odd contributions.
During the collision process, a local spin polarization appears as
was first observed in TDHF calculations~\cite{Mar06,Iwa11}. It is
clearly seen that the spins of participant and spectator nucleons
are more likely to be up and down, respectively, in the most
compressed stage of the reaction. The spin polarization follows the
direction of the vector $\nabla \times \vec{j}$ rather than that of
$\vec{p} \times \nabla \rho$ since the latter has a smaller
magnitude although it points to the opposite direction of the
former. In the later stage, however, the spin polarization becomes
weaker because of NN scatterings and other spin mixing effects,
especially for participant nucleons in the high-density region.

\begin{figure}[t]
\includegraphics[scale=0.6]{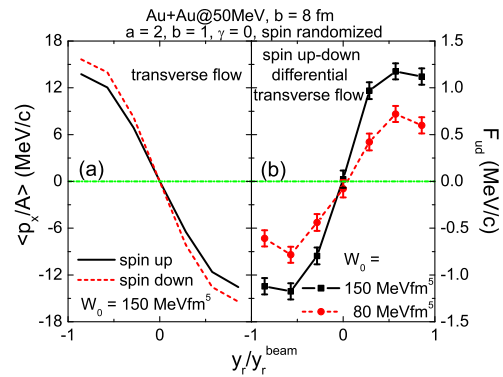}
\caption{(color online) Transverse flow of spin-up nucleons and
spin-down nucleons (a) and spin up-down differential flows using
different $W_0$ (b) for the same reaction as in
Fig.~\ref{den}.}\label{Fud1}
\end{figure}

As shown in the equations of motion, the spin-orbit coupling also
affects the nucleon momentum and spatial distributions besides the
spin polarization. Nucleon transverse collective flow, measured by
using the average transverse momentum $<p_x(y_r)>$ in the reaction
plane versus rapidity $y_r$, is one of the best known observable for
revealing effects of density gradients in nuclear reactions
\cite{Ber88,Dan02a,Dan85}. Since spin-up and spin-down nucleons with
the same momentum experienced opposite spin-related potentials
during the whole collision process, we expect the difference in
transverse flow of spin-up and spin-down nucleons to be sensitive to
the spin-orbit coupling while other effects will be largely canceled
out. To test this idea, we first compare in the left panel of
Fig.~\ref{Fud1} the transverse flows of spin-up and spin-down
nucleons. It is seen that the transverse flow of spin-up nucleons is
smaller than that of spin-down ones. This can be understood by
looking at the x component of the density gradient and the y
component of the curl of the momentum density shown in the third and
the fourth row of Fig.~\ref{den}. By examining the time evolution,
we found that the effects of the spin-orbit interaction on the
transverse flow during the first $40$ fm/c of the collision are
mostly washed out due to violent interactions. The spin-dependent
transverse flow is mainly determined by the dynamics afterwards. As
the projectile (target) is still moving in the +z (-z) direction,
the participant nucleons from the projectile (target) with negative
(positive) $(\nabla \rho)_x$ give a more repulsive/attractive
spin-orbit potential [$\nabla \rho \cdot (\vec{p} \times
\vec{\sigma})>/<0$] for spin-up/down nucleons. This leads to a
larger transverse flow for spin-up nucleons than spin-down ones. On
the other hand, the time-odd term contributes exactly in the
opposite direction and is stronger than the time-even term. The
combined effects therefore lead to a smaller (larger) transverse
flow for spin-up (spin-down) nucleons.

To extract more accurately information about the spin-related
potentials without much hinderance of spin-independent potentials,
we investigate next the spin up-down differential transverse flow
\begin{equation}
F_{ud}(y_r) = \frac{1}{N(y_r)} \sum_{i=1}^{N(y_r)} \sigma_i (p_x)_i,
\end{equation}
where $N(y_r)$ is the number of nucleons with rapidity $y_r$, and
$\sigma_i$ is $1 (-1)$ for spin-up (spin-down) nucleons. Similar to
the neutron-proton differential transverse flow for probing the
symmetry potential~\cite{Li00}, the spin up-down differential
transverse flow maximizes the effects of the opposite spin-related
potentials for spin-up and spin-down nucleons while canceling out
largely spin-independent contributions. Indeed, the spin up-down
transverse flow is a sensitive probe of the spin-orbit coupling
strength $W_0$. As an example, shown in the right panel of
Fig.~\ref{Fud1} is a comparison of the spin up-down differential
transverse flows obtained using the two limiting values of $W_0$
used in the literature. To be conservative, in this example we have
used the randomized spin assignment after each NN scattering which
is the worst scenario for revealing effects of the spin-orbit
potential. Fortunately, even in this case, a 47\% increase in $W_0$
leads to an approximately 40\% higher up-down differential flow far
beyond the statistical errors in the calculation.

\begin{figure}[t]
\includegraphics[scale=0.6]{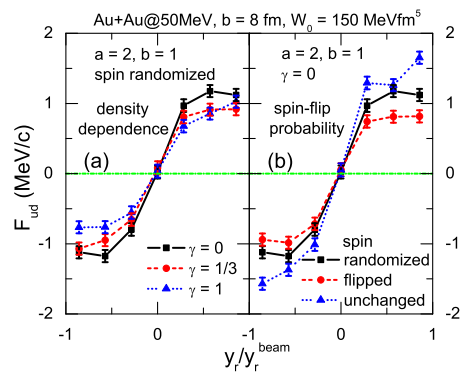}
\caption{(color online) Spin up-down differential transverse flows
using different density dependence of the spin-orbit coupling
coefficients (a) and different treatments of spin after NN
scatterings (b) for the same reaction as in
Fig.~\ref{den}.}\label{Fud2}
\end{figure}

The density dependence of the spin-orbit coupling, which was tested
earlier, see, e.g., Ref.~\cite{Pea94}, is still almost completely
unknown, and this has recently motivated more new
experiments~\cite{Sorlin12}. To investigate effects of the density
dependence of the spin-orbit coupling on the spin up-down
differential flow, shown in Panel (a) of Fig.~\ref{Fud2} are the
results obtained by varying only the $\gamma$ parameter. It is seen
that the spin up-down differential flow is larger for a weaker
density dependence of the spin-orbit coupling if its strength at
saturation density is fixed.

\begin{figure}[t]
\includegraphics[scale=0.6]{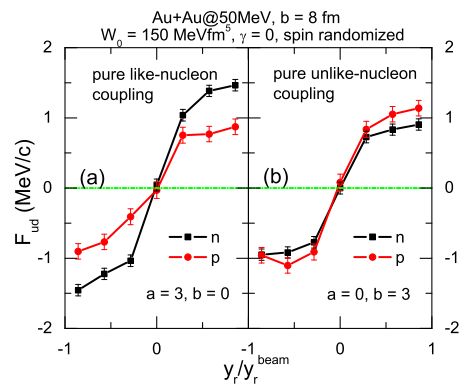}
\caption{(color online) Spin up-down differential transverse flows
for neutrons and protons from pure like-nucleon coupling (a) and
pure unlike-nucleon coupling (b) for the same reaction as in
Fig.~\ref{den}.}\label{Fud3}
\end{figure}

The isospin dependence of the spin-orbit coupling is another
interesting issue especially relevant for understanding the
structure of rare isotopes and the synthesis of superheavy elements.
To evaluate potential applications of our approach in further
constraining the isospin dependence of the spin-orbit interaction,
we next compare the spin up-down differential flows for neutrons and
protons using the pure like-nucleon coupling ($a=3$ and $b=0$) and
pure unlike-nucleon coupling ($a=0$ and $b=3$) in Panel (a) and
Panel (b) of Fig.~\ref{Fud3}, respectively. As the system considered
is globally neutron-rich and $\nabla \rho_n$ and $\nabla \times
\vec{j}_n$ are generally larger than $\nabla \rho_p$ and $\nabla
\times \vec{j}_p$, respectively, the pure like (unlike)-nucleon
coupling leads to an appreciably larger (smaller) spin up-down
differential flow for neutrons than for protons. Moreover, the
unlike-nucleon coupling generally reduces slightly the overall
strength of the spin-related potentials and thus the spin up-down
differential flow. Of course, more neutron-rich systems will be
better for probing the isospin dependence of the spin-orbit coupling
using the double differential flow between spin up-down neutrons and
protons.

Finally, what are the effects of the possible spin flip in NN
scatterings on the spin up-down differential flow? We answer this
question quantitatively by using the results shown in the right
panel of Fig.~\ref{Fud2}. Due to the lack of knowledge about the
energy and isospin dependence of the spin-flip probability for
in-medium nucleon-nucleon scatterings, we compare results obtained
by using the following three choices for setting the final spins of
colliding nucleons after each NN scattering: (1) flipped, (2)
randomized, or (3) unchanged, effectively varying the spin-flip
probability from large to small. It is seen that the spin up-down
differential transverse flow decreases with increasing spin-flip
probability as one expects. Moreover, it is very encouraging to see
that the spin up-down differential flow is still considerable even
if a 100$\%$ spin-flip probability is assumed, further proving the
validity of using it as a probe of the spin-orbit coupling.

In summary, the spin degree of freedom and the spin-related
potentials are incorporated for the first time in an
isospin-dependent transport model providing a useful new tool for
investigating the spin-isospin dynamics of heavy-ion collisions at
intermediate energies, such as the development of local spin
polarization in these reactions. The nucleon spin up-down
differential transverse flow is shown to be a sensitive probe of the
in-medium spin-orbit interaction. Comparisons with future
experiments will allow us to determine the density and isospin
dependence of the in-medium spin-orbit coupling that has significant
ramifications in both nuclear physics and astrophysics.

We thank Y. Iwata and K. Asahi for useful communications
and W. G. Newton, F. J. Fattoyev, and Z. T. Liang for
helpful discussions. This work was supported in part by the US
National Science Foundation grants PHY-0757839 and PHY-1068022, the
National Aeronautics and Space Administration under grant NNX11AC41G
issued through the Science Mission Directorate,  the CUSTIPEN
(China-U.S. Theory Institute for Physics with Exotic Nuclei) under
DOE grant number DE-FG02-13ER42025, and the '100-talent plan' of
Shanghai Institute of Applied Physics under grant Y290061011 from
the Chinese Academy of Sciences.

\end{document}